\documentstyle[epsf,epsfig,12pt]{article}  
\begin{document}

\newcommand{\be}{\begin{equation}}

\begin{titlepage}

\pagenumbering{arabic}
\vspace*{-1.cm}
\begin{tabular*}{15.cm}{l@{\extracolsep{\fill}}r}
&
\hfill \bf{DEMO-HEP 98/04}
\\
& 
\hfill November 1998
\\
\end{tabular*}
\vspace*{2.cm}
\begin{center}
\Large 
{\bf Extended Modified Observable Technique\\
for a Multi-Parametric Trilinear Gauge Coupling \\
Estimation at LEP II}\\

\vspace*{2.cm}
\normalsize { 
{\bf G. K. Fanourakis, D. Fassouliotis, A. Leisos, \\
N. Mastroyiannopoulos and  S. E. Tzamarias} \\
{\footnotesize Institute of Nuclear Physics - N.C.S.R. Demokritos\\
               15310  - Aghia Paraskevi  - Attiki - Greece}\\
}
\end{center}
\vspace{\fill}
\begin{abstract}
This paper describes the extension of the Modified Observables technique in estimating
simultaneously more than one Trilinear Gauge Couplings. The optimal properties,
unbiasedness and consistent error estimation of this method are demonstrated by
Monte Carlo experimentation using $\ell \nu jj $ four-fermion final state topologies.
Emphasis is given in the determination of the expected sensitivities in estimating
the $\lambda_{\gamma} - \Delta g_{1}^{z}$ and $\Delta k_{\gamma} - \Delta g_{1}^{z}$
pair of couplings with data from the 183 GeV LEPII run.
\noindent
\end{abstract}
\vspace{\fill}

\end{titlepage}

\pagebreak

\begin{titlepage}
\mbox{}
\end{titlepage}

\pagebreak

\setcounter{page}{1}    


\section{Introduction}

It has been shown \cite{oo}, that by expanding the probability distribution
function (p.d.f.)  and keeping only linear terms with respect to the Trilinear
Gauge Couplings (TGC's), one
can build estimators (the Optimal Observables) which are linear functions of 
the couplings around the expansion point. Furthermore, this linear dependence 
can be easily evaluated by theory. An efficient estimation of the couplings 
can be performed by inverting these linear relations. 
Such an estimation has the same accuracy as the unbinned maximum 
likelihood technique.

The method of the Optimal Observables has been extended \cite{ourmo} to 
incorporate the influence of the detector effects to the measurement of the
kinematical vectors. An iterative procedure has been also introduced to ensure
the consistency and optimality of the technique, independent of the choice of
the parametric expansion point. In the same paper, the optimal properties of 
this (Modified Observables) method have been demonstrated 
for one coupling fits to the 172 GeV LEPII data.\\
In the  meanwhile larger data samples are available from the 183 GeV LEPII
run and the application of the Modified Observable technique to a 
simultaneous estimation of two couplings is very relevant. 

This paper concentrates on the simultaneous estimation of two TGC's by
employing phenomenological models \cite{yreport} where two couplings could
deviate freely from their Standard Model (S.M.) values whilst  certain constraints
are imposed on the other couplings.
This paper is dealing with 
WW events produced in $e^{+}e^{-}$ annihilation,
 where one of the W's decays leptonically whilst the other decays in two jets. 
A large sample of 60000 $WW\rightarrow \ell \nu q\bar{q}$ 
Monte Carlo (M.C.) events was  used to evaluate cross sections and other 
statistics, as well as their dependence on the coupling values by the 
M.C. reweighting procedure \cite{rew}. 
These events have been produced either by PYTHIA \cite{pythia} 
(employing only the CC03 production diagrams) or by EXCALIBUR \cite{exca} 
(full 4-fermion production) 
at different coupling values and they have undergone full detector
simulation by the DELSIM \cite{delsim} simulation programme. 
Moreover these events have  been
reconstructed and selected by the same analysis algorithms as the real data
 \cite{del183} \cite{our183} accumulated with the DELPHI \cite{delphi}
 detector. 
The background contamination has been simulated by the 
production of the physics channels \cite{del183} \cite{our183} \cite{delww} 
which produce final state topologies indistinguishable from the signal WW events. 

This paper is organised as follows: 
the statistical technique and its asymptotic properties are described in 
Section 2, whilst numerical results obtained by  M.C.
 experimentation are presented
in Section 3. Finally, Section 4 contains the comparison with other 
techniques and the conclusions.

\section{ Modified Observables in Multi-Parametric Fits}

The present study is  focusing on two parameter 
(TGC's) estimations but this 
analysis can be extended to any number of parameters in a straight forward way.

The probability distribution function, with respect to the observed
kinematical vector
 $\vec{\Omega}$, is expressed  \cite{ourmo} \cite{yreport} as a function of the 
two couplings $\alpha_1$ and $\alpha_2$ as

\begin{eqnarray}
P(\vec{\Omega};\alpha_1 ,\alpha_2) = 
 \int \frac
{d\sigma (\vec{V};\alpha_{1},\alpha_{2})/d \vec{V}}
{\sigma_{tot}(\alpha_{1}, \alpha_{2})}
\cdot \epsilon(\vec V) \cdot R(\vec V ;\vec{\Omega}) \cdot d\vec V
\nonumber \\
= \int \frac
{\sum_{i=0}^{2} \sum_{j=0}^{2-i} c_{ij}(\vec V)\alpha_{1}^{i} \alpha_{2}^{j}}
{\sum_{i=0}^{2} \sum_{j=0}^{2-i} S_{ij}\alpha_{1}^{i} \alpha_{2}^{j}}
\cdot \epsilon(\vec V) \cdot R(\vec V ;\vec{\Omega}) \cdot d\vec V
\label{eq:1}
\end{eqnarray}
where

\begin{description}
\item[$\vec V$] is the true kinematical vector which describes the events 
\item[$\epsilon(\vec V)$] is the efficiency of observing an event produced at
$\vec V$
\item[$R(\vec V;\vec{\Omega})$] is the resolution function, i.e. the
probability the true kinematical vector $\vec{V}$ to be measured as
$\vec{\Omega}$
\item[$\frac {d\sigma (\vec{V};\alpha_{1},\alpha_{2})} {d \vec{V}}$] is the
differential cross-section
\item[$\sigma_{tot}(\alpha_{1}, \alpha_{2})$] is the total cross-section and
\item[ $S_{ij} = \int c_{ij}(\vec V) d\vec V$.] 
\end{description}

In \cite{ourmo}, it has been shown that the Optimal Observables including detector 
effects, in the neighbourhood 
of the parametric point $\{ \alpha_1^0,\alpha_2^0 \}$, 
are defined as the mean values of the following quantities:

\begin{eqnarray}
z_1(\vec{\Omega};\alpha_1^0,\alpha_2^0) = \frac 
{\int y_{01}(\vec V;\alpha_1^0,\alpha_2^0) \epsilon(\vec V) 
R(\vec V;\vec{\Omega})d\vec V}
{\int y_{00}(\vec V;\alpha_1^0,\alpha_2^0) \epsilon(\vec V) 
R(\vec V;\vec{\Omega})d\vec V} \nonumber \\
z_2(\vec{\Omega};\alpha_1^0,\alpha_2^0) = \frac 
{\int y_{10}(\vec V;\alpha_1^0,\alpha_2^0) \epsilon(\vec V) 
R(\vec V;\vec{\Omega})d\vec V}
{\int y_{00}(\vec V;\alpha_1^0,\alpha_2^0) \epsilon(\vec V) 
R(\vec V;\vec{\Omega})d\vec V}
\label{eq:2}
\end{eqnarray}

where the functions $y_{\kappa,\lambda}(\vec{V};\alpha_1^0,\alpha_2^0)$ are
expressed in terms of the differential cross section coefficients as:
\begin{eqnarray}
y_{00}(\vec{V};\alpha_1^0,\alpha_2^0) &=& c_{00}(\vec{V})+c_{01}
(\vec{V})\alpha_1^0 +c_{10}(\vec{V})\alpha_2^0+c_{20}(\vec{V})\alpha_1^{_{0}2}
 + 
c_{02}(\vec{V})\alpha_2^{_{0}2}+ c_{11}(\vec{V})\alpha_1^0 \alpha_2^0 
\nonumber \\
y_{10}(\vec{V};\alpha_1^0,\alpha_2^0) &=& c_{10}(\vec{V})+
2c_{20}(\vec{V})\alpha_1^0 + c_{11}(\vec{V})\alpha_2^0  \label{eq:3}\\ 
y_{01}(\vec{V};\alpha_1^0,\alpha_2^0)&=&c_{01}(\vec{V})+2c_{02}(\vec{V})
\alpha_2^0 +
c_{11}(\vec{V})\alpha_1^0 \nonumber
\end{eqnarray}

It has also been shown that the Optimal Observables 
are linear functions of the couplings $\alpha_1$ and $\alpha_2$ in the 
neighbourhood of ($\alpha_1^0,\alpha_2^0$) 
\footnote{This is easily proven
by expanding (\ref{eq:1}) in a Taylor series around \{ $\alpha_1^0,\alpha_2^0$\}
and evaluating the mean values of (\ref{eq:2}) ignoring higher than first
order in $\alpha_1 - \alpha_1^0$ and $\alpha_2 - \alpha_2^0$ terms.} 
i.e.:

\begin{eqnarray}
\int z_k(\vec{\Omega};\alpha_1^0,\alpha_2^0)
P(\vec{\Omega};\alpha_1,\alpha_2)d\vec{\Omega} =
&& \int z_k(\vec{\Omega};\alpha_1^0,\alpha_2^0)
 P(\vec{\Omega};\alpha_1^0,\alpha_2^0)d\vec{\Omega}+  \nonumber \\
&& \sum_{i=1}^{2} [ \int z_i(\vec{\Omega};\alpha_1^0,\alpha_2^0)
z_k(\vec{\Omega};\alpha_1^0,\alpha_2^0) 
P(\vec{\Omega};\alpha_1^0,\alpha_2^0)d\vec{\Omega}- \nonumber \\
&& ( \int z_i(\vec{\Omega};\alpha_1^0,\alpha_2^0)
P(\vec{\Omega};\alpha_1^0,\alpha_2^0)d\vec{\Omega} \cdot  \nonumber \\
&&  \int z_k(\vec{\Omega};\alpha_1^0,\alpha_2^0)
P(\vec{\Omega};\alpha_1^0,\alpha_2^0)d\vec{\Omega} )]
\cdot (\alpha_{i} - \alpha_{i}^{0}) 
\label{eq:4}
\end{eqnarray}
where k=1,2


Thus, given a set of N experimentally measured vectors $\vec{\Omega}_n$
 (n=1,\ldots ,N) the left hand side of (\ref{eq:4}) can be approximated as:
\begin{equation}
\int z_k(\vec{\Omega};\alpha_1^0,\alpha_2^0)\vec
P(\vec{\Omega};\alpha_1,\alpha_2)d\vec{\Omega} \simeq
\frac {1}{N} \sum_{n=1}^{N}z_k(\vec{\Omega}_n;\alpha_1^0,\alpha_2^0)
\label{eq:4a}
\end{equation}
The right hand side  of (\ref{eq:4}) can be calculated 
using the theoretical 
expression of the cross section as a function of the couplings,
provided that the resolution and 
efficiency functions can be parametrized analytically. Then, a simple inversion
of the linear system of equations (\ref{eq:4}) results to an estimation
of the coupling values with the same sensitivity  as with the maximum 
likelihood technique.\\
In practice, neither the efficiency nor the resolution
function can be parametrized analytically. However, it has been shown that 
a very succesfull approximative way of using the basic concepts of the 
Optimal Observables in one TGC parameter estimations \cite{ourmo} exists.
That is the Modified Observable technique, which in this paper is extended
to more than one TGC simultaneous estimations.

Following the same steps as in \cite{ourmo}, the functional forms
\footnote{$z_{1,2}(\vec{\Omega};\alpha_1^0,\alpha_2^0)$ are defined in 
\cite{ourmo} as the mean values of the quantities 
$\frac {y_{01,10}(\vec V;\alpha_1^0,\alpha_2^0)}
       {y_{00}(\vec V;\alpha_1^0,\alpha_2^0)}$ corresponding
to kinematical vectors $\vec V$ produced with coupling values 
$\alpha_1^0$ and $\alpha_2^0$ and being observed in the phase space
element $\vec{\Omega}\cdot d\vec{\Omega}$.} 
of $z_{1}(\vec{\Omega};\alpha_1^0,\alpha_2^0)$ and 
$z_{2}(\vec{\Omega};\alpha_1^0,\alpha_2^0)$
in (\ref{eq:2}) are approximated as:
\begin{equation}
z_{1,2}(\vec{\Omega};\alpha_1^0,\alpha_2^0) \simeq 
\frac {y_{01,10}(\vec{\Omega};\alpha_1^0,\alpha_2^0)}
      {y_{00}(\vec{\Omega};\alpha_1^0,\alpha_2^0)} 
\label{eq:5}
\end{equation}
These are very good approximations, as indicated in figure
\ref{compare} where  the mean values
of the quantities $\frac {y_{01,10}(\vec V;\alpha_1^0,\alpha_2^0)}
      {y_{00}(\vec V;\alpha_1^0,\alpha_2^0)}$ 
are compared with the quantities
$\frac {y_{01,10}(\vec{\Omega};\alpha_1^0,\alpha_2^0)}
      {y_{00}(\vec{\Omega};\alpha_1^0,\alpha_2^0)}$, for 
several expansion points  $\alpha_1^0$ and $\alpha_2^0$.
These mean values have been 
evaluated by using M.C. events produced with 
coupling values $\alpha_1^0$ and $\alpha_2^0$ and being observed
with kinematical vector $\vec{\Omega}$ corresponding to a bin of
$\frac {y_{01,10}(\vec{\Omega};\alpha_1^0,\alpha_2^0)}
      {y_{00}(\vec{\Omega};\alpha_1^0,\alpha_2^0)}$. \\
The functional form of $z_{1,2}(\vec{\Omega};\alpha_1^0,\alpha_2^0)$ in (\ref{eq:5})
is independent of phase space and other multiplicative  (e.g. Initial State Radiation)
factors and it was calculated by using the ERATO \cite{erato} four-fermion matrix
element package by folding the kinematical information corresponding to the two
hadronic jets.

Instead of calculating the terms of the right hand side of (\ref{eq:4}),
the dependence of the mean values of 
$z_{1,2}(\vec{\Omega};\alpha_1^0,\alpha_2^0)$ (in the following called
Modified Observables ) on the production values of the couplings
has been evaluated by reweighted M.C. \cite{rew} integration.\\ 
Figure  \ref{surfaces} shows the surfaces 
$f_{1}(\alpha_1,\alpha_2;\alpha_1^0,\alpha_2^0)$ and
$f_{2}(\alpha_1,\alpha_2;\alpha_1^0,\alpha_2^0)$
(in the following called calibration surfaces),
which express the dependence
of the product of each Modified Observable
 with the number of expected events for
luminosity of 50.23 $pb^{-1}$, as a function of the coupling values, for
three initial parametric points. These products (instead of the Modified
Observables themselves) are going to be used as estimators 
of the couplings, gaining more efficiency by including  the
extra information of the total number of the observed events
 \cite{oo}.\\
The couplings are estimated by 
 comparing the calibration surfaces  to the experimental measurements,
that is to the products of the measured values of the Modified Observables 
with the number of observed events, which are simply expressed as:
\begin{eqnarray}
d_1(\alpha_1^0,\alpha_2^0) = 
\sum_{i=1}^{N} z_1(\vec{\Omega}_{i};\alpha_1^0,\alpha_2^0) \nonumber \\
d_2(\alpha_1^0,\alpha_2^0) = 
\sum_{i=1}^{N} z_2(\vec{\Omega}_{i};\alpha_1^0,\alpha_2^0) \label{eq:6bb}
\end{eqnarray}
Such  comparisons are shown in figures 
(\ref{surf2_00}) and (\ref{surf2_0m10p1}) between a large independent
set of M.C. events used as a data sample and three pairs of calibration
surfaces ($f_{1}(\alpha_1,\alpha_2;\alpha_1^0,\alpha_2^0)$ and
$f_{2}(\alpha_1,\alpha_2;\alpha_1^0,\alpha_2^0)$)
evaluated at three different expansion points 
$\{\alpha_1^0,\alpha_2^0\}$. 
In these figures the intersections of the calibration surfaces with the planes
defined by the experimental measurements, 
$d_1(\alpha_1^0,\alpha_2^0)$ and $d_2(\alpha_1^0,\alpha_2^0)$, are also shown.
It is worth noticing that the estimation, which is the common point
of the pair of lines in figures 
\ref{surf2_00}c, \ref{surf2_0m10p1}c and \ref{surf2_0m10p1}f, 
is independent from the expansion point. 
This fact reflects one of the basic properties of the technique to be 
globally unbiased. \\
However, the evaluation of the estimation confidence intervals is more 
complicated, due to the statistical correlations between 
the calibration surfaces as well as  between the measured quantities
$d_1(\alpha_1^0,\alpha_2^0)$ and $d_2(\alpha_1^0,\alpha_2^0)$.\\ 
The covariant matrices
$M(\alpha_1,\alpha_2;\alpha_1^0,\alpha_2^0)$ 
(expressing the statistical accuracy of the calibration surface evaluation at
the expansion point \{ $(\alpha_1^0,\alpha_2^0)$ \} ) and
$V(\alpha_1^0,\alpha_2^0)$ (which is the covariant matrix corresponding to the 
measured quantities 
$d_1(\alpha_1^0,\alpha_2^0)$ and $d_2(\alpha_1^0,\alpha_2^0)$) are calculated 
 from the kinematical vectors of the reweighted M.C. and real
events respectively.\\
Then, assuming gaussian errors, the probability that the selected 
 event sample supports coupling values equal to $\alpha_1$ and 
$\alpha_2$,  is given by the Likelihood function:

\begin{eqnarray}
L=\frac {1}{2\pi \left| W \right|} \cdot exp  [ -\frac {1}{2} 
\left( \vec D(\alpha_1^0,\alpha_2^0)-
\vec F(\alpha_1,\alpha_2;\alpha_1^0,\alpha_2^0) \right) ^{T}
\cdot \nonumber \\
W^{-1} \cdot 
 \left( \vec D(\alpha_1^0,\alpha_2^0)-
\vec F(\alpha_1,\alpha_2;\alpha_1^0,\alpha_2^0) \right) ] \label{eq:88}
\end{eqnarray}
where the vector $\vec D(\alpha_1^0,\alpha_2^0)$, the
vector calibration function 
$\vec F(\alpha_1,\alpha_2;\alpha_1^0,\alpha_2^0)$ and
the covariant matrix $W(\alpha_1,\alpha_2;\alpha_1^0,\alpha_2^0)$
are defined as follows:

\begin{equation}
\vec D(\alpha_1^0,\alpha_2^0)= 
\left( \begin{array}{l}
d_1(\alpha_1^0,\alpha_2^0) \\
d_1(\alpha_1^0,\alpha_2^0) 
\end{array} \right) \label{eq:6}
\end{equation} 

\begin{equation}
\vec F(\alpha_1,\alpha_2;\alpha_1^0,\alpha_2^0)=
\left( \begin{array}{l}
f_1(\alpha_1,\alpha_2;\alpha_1^0,\alpha_2^0) \\
f_2(\alpha_1,\alpha_2;\alpha_1^0,\alpha_2^0) 
\end{array} \right) \label{eq:7}
\end{equation} 

\begin{equation}
W(\alpha_1,\alpha_2;\alpha_1^0,\alpha_2^0) = 
M(\alpha_1,\alpha_2;\alpha_1^0,\alpha_2^0) +
V(\alpha_1^0,\alpha_2^0)
\end{equation}

Maximization of (\ref{eq:88}), with respect to 
 $\alpha_1$ and $\alpha_2$, provides
the estimation of the coupling values, whilst the confidence intervals are 
evaluated\footnote{
$-(log L_{max} - 1.205)$ for 70\% confidence intervals}
by the asymptotic gaussian properties of the estimation 
distribution \cite{eadie}.\\
A set of M.C. events produced with Standard Model coupling
values ( 6000 events at  $\{\lambda_{\gamma}=0,\Delta g_{1}^{z}=0\}$),
was used as data sample to demonstrate the asymptotic properties 
of such estimations. The $\lambda_{\gamma}, \Delta g_{1}^{z}$ 
couplings were simultaneously estimated by maximizing the likelihood 
function of (\ref{eq:88}) and the estimated
coupling values are shown as functions of the expansion point in
figure (\ref{converge}). The fact
that the estimations are close (within the statistical errors) to the true
coupling values, for the whole region of the expansion points,  emphasizes
the optimal properties of the method.
However, the optimal estimated error is achieved \cite{oo} at 
 expansion points close to the estimated values, where the
linear dependance of the Optimal Variables holds.
This is shown in  figure \ref{converge}c where 
three 70\% confidence limit contours corresponding to different expansion points
are presented for comparison.
Obviously
the optimum estimated sensitivity is achieved in the case where 
$\alpha_1^0 = \hat{\alpha_1}$ and $\alpha_2^0 = \hat{\alpha_2}$, where
$\hat{\alpha_1}$ and $\hat{\alpha_2}$ are the estimated values. 
The fact that the above condition also guarantees a correct error estimation,
is demonstrated in the next section by Monte Carlo experimentation.

\section{Numerical results }
A series of M.C. experiments were used to demonstrate the optimal properties
of the Modified Observable technique when  two
TGC's are simultaneously estimated by fitting finite statistical samples.

Fully reconstructed four fermion EXCALIBUR events,  produced with S.M. 
coupling values, were mixed with background events to form data 
sets corresponding to the luminosity of 50.23 $pb^{-1}$ accumulated  by the
DELPHI detector at $\sqrt{s}~\simeq~183$ GeV.
Each set consisted of 82, 101 and 39 events in average with an 
electron, muon and tau lepton in the final state, respectively. 
The average background contribution to each of the above subsets were 
8.0, 1.4 and 8.3 events. The specific event multiplicity of each data 
set was chosen to follow poissonian distributions.
Another set of fully four fermion and background reconstructed events, 
produced and selected as described in Section 1, was used 
to calculate  cross sections and probabilties as well as  their 
dependence on the TGC's  by reweighted Monte Carlo integration.
In fitting the data sets, the ($\lambda_{\gamma},\Delta g^{Z}_{1}$) 
and the  ($\Delta k_{\gamma},\Delta g^{Z}_{1}$)
TGC schemes were used \cite{yreport} and 
a simultaneous estimation of the free couplings was performed.\\
In order to take into account the differences in the production dynamics, 
the selection
efficiencies and   
the background contamination between the final states ($\ell \nu j j$, 
$\ell= \mu,e,\tau$) the measured vector 
$\vec D(\alpha_1^0,\alpha_2^0)$ was defined as follows:
\begin{equation}
\vec D(\alpha_1^0,\alpha_2^0)= 
\left( \begin{array}{l}
\sum_{\ell=1}^{3} (d_{1}^{\ell}(\alpha_1^0,\alpha_2^0) -B_{1}^{\ell}) \\
\sum_{\ell=1}^{3} (d_{2}^{\ell}(\alpha_1^0,\alpha_2^0) -B_{2}^{\ell})
\end{array} \right) \label{eq:6aa}
\end{equation} 
Where $\ell=1,2,3$ stands for the three lepton tags whilst
$B_{1}^{\ell}$, $B_{2}^{\ell}$ denotes the expected contribution of the
background  events to the measurement.\\
Similarly the calibration surface vector was defined as:
\begin{equation}
\vec F(\alpha_1,\alpha_2;\alpha_1^0,\alpha_2^0)=
\left( \begin{array}{l}
\sum_{\ell=1}^{3} f_{1}^{\ell}(\alpha_1,\alpha_2;\alpha_1^0,\alpha_2^0) \\
\sum_{\ell=1}^{3} f_{2}^{\ell}(\alpha_1,\alpha_2;\alpha_1^0,\alpha_2^0) 
\end{array} \right) \label{eq:7aa}
\end{equation} 

\pagebreak

The asymptotic property of the log likelihood ratio 
\cite{eadie} was used  
to demonstrate the unbiasedness of  the proposed techniques.
That is, the  $\chi^2$ 
(n.d.f.=2) probability of obtaining the specific value of $\lambda$, where 
\begin{equation}
\lambda = -2\cdot\log \frac{L(\alpha_{1}^{true},\alpha_{2}^{true})}
{L(\hat{\alpha}_{1},\hat{\alpha}_{2})}
\label{eq:22b}
\end{equation}
in fits of the data sets should follow an equiprobable distribution.\\
Furthermore, the consistency in evaluating the error matrix of the
estimated couplings  
($\hat{\cal {E}}$) in every fit, is checked 
by using the other asymptotic property \cite{eadie} of the 
likelihood estimations 
to be gaussian distributed around the true parameter values. 
Thus for an unbiased estimation of the central values and for a correct 
error matrix evaluation the quantity $\delta$:
\begin{equation}
\delta=
\left( \begin{array}{l}
\hat{a_{1}}-a_{1}^{true} \\
\hat{a_{2}}-a_{2}^{true} 
\end{array} \right) \cdot \hat{\cal{E}} \cdot 
\left(\begin{array}{clcr}
\hat{a_{1}}-a_{1}^{true} &
\hat{a_{2}}-a_{2}^{true}
\end{array}\right) \label{eq:23}
\end{equation}
should follow a $\chi^2$(n.d.f.=2) distribution.
This property is demonstrated by presenting the $\chi^2$(n.d.f.=2) 
probabilities to obtain specific $\delta$ values in fitting
the data sets.\\
The above  tests of $\lambda$  and $\delta$ $\chi^2$-probability 
distributions can be considered as  
extensions of the sampling and pull distribution tests respectively, 
commonly used in one parameter fits.

Due to the limited number of the available M.C. events, only sixty 
independent data sets could be constructed. 
Although the number of the data sets  is enough to show the 
optimal properties of the proposed technique,  the 
bootstrap procedure 
\footnote
{The bootstrap procedure advocates that one can select randomly 
${\cal{N}}$ events to form a set from a pool of ${\cal{K}}$ 
available events, and repeat the random selection to construct many 
bootstrapped sets. The distribution of 
statistics evaluated from each of the bootstrapped set approximates 
well the true distribution, as long as ${\cal{K}}$ is big enough 
compared to ${\cal{N}}$.} 
\cite{boot} has been also used to construct a large number of 
semicorrelated data sets. 
 
Results of estimating the  ($\lambda_{\gamma},\Delta g^{Z}_{1}$) and 
($\Delta k_{\gamma},\Delta g^{Z}_{1}$)  couplings with 
the Modified Observable technique are shown in figure \ref{chi}. 
In both TGC schemes the optimal properties of the technique in 
estimating central values and error matrices are obvious. 
Specifically the sixty completely uncorrelated samples 
produce $\chi^{2}(n.d.f.=2)$ probabilities distributed 
with mean values close to 0.5 and root mean squares close to $1/\sqrt{12}$, 
whilst the equiprobable behaviour of the
 $\chi^{2}(n.d.f.=2)$ probabililty values obtained by fitting the 
bootstrapped samples is striking.

The $\chi^{2}$ behaviour of the $\lambda$ and $\delta$  
quantities is further used to quantify the sensitivity of this technique. 
Indeed such property \cite{eadie}
ensures that the estimated values $\{\hat{\alpha_{1}},\hat{\alpha_{2}}\}$ 
follow a two dimensional gaussian distribution with a covariant matrix 
which characterises
the average sensitivity in estimating the couplings. The covariant
matrix elements  (i.e.
 the variances and correlations of the couplings estimations) are found 
by fitting a 2-dim 
gaussian to the estimated coupling values from the 60 independent sets.
These average sensitivities are summarized in Tables 1 and 2 for 
($\lambda_{\gamma},\Delta g^{Z}_{1}$) and 
($\Delta k_{\gamma},\Delta g^{Z}_{1}$) estimations.\\
The same uncorrelated M.C. sets of events were treated as if they have been
collected by a ``perfect'' detector and the two pairs of couplings 
were estimated
by an unbinned extended likelihood fit
 as well as by the Modified Observable technique
\footnote
{ The true kinematical vector $\vec{V}$ of each
event was used to calculate the matrix element and the
calibration surfaces. In the following, when $\vec{V}$
is used, the methods and their results will be named as
``perfect''.}. 
The average sensitivities obtained from these 
estimations (``perfect'' detector extended unbinned likelihood 
and ``perfect'' detector Modified Observables)
are also shown for comparison in Tables 1 and 2 where the equivalence of the
Modified Observables to the likelihood fit is obvious. 
The loss of sensitivity in the case of a realistic detector is a 
natural consequence of the loss of information due to the imperfect
experimental resolution. However, the consistent inclusion of the 
detector effects in the realistic case guaranties consistent central value 
and confidence interval estimation. It is also worth noticing that 
in the realistic case, the evaluated errors and correlations in
every individual Modified Observable estimation are gaussian 
distributed with means very close to the average sensitivities, as it is 
shown in figure \ref{errors}.

As a last point, figure \ref{dkg} shows the  sampling, the pull and the error 
distributions of a single coupling ($\Delta k_{\gamma}$) estimation. 
Similar estimations of the same coupling \cite{ourmo}, 
using 172 GeV data samples, have been found to exhibit non gaussian tails.
However, 
it was advocated that with small event
samples, where the statistical error is large compared
to the linear part of the calibration curves, the evaluated error 
from the fits is expected to underestimate the sensitivity of the technique.
Obviously such pathologies are absent when the relative statistical error is smaller,
as in the case of the data sample accumulated at 183 GeV.

\section{Conclusion}

In this paper the Modified Observables technique \cite{ourmo} was 
generalized in order to be applied for a simultaneous estimation of two 
couplings by deploying the appropriate TGC scheme \cite{yreport}. 
The technique,including the detector effects and the background contribution, 
was demonstrated to be asymptotically a consistent  estimator. 
This consistency was
also shown to be independent of the initial expansion  values.
However the optimal sensitivity is achieved
for  expansion points close to the estimated values of the couplings.\\
The properties of the technique, when fitting finite size event samples, were
investigated by M.C. experimentation. Sets of M.C. events, of the same size
as the data samples accumulated by each of the LEP experiments at $\sqrt{s}~\simeq~183$ GeV, 
were fitted to estimate the $\{\lambda_{\gamma},\Delta g_{1}^{z}\}$
and $\{\Delta \kappa_{\gamma},\Delta g_{1}^{z}\}$ pairs of couplings. 
The distributions of these estimations demostrated the optimal 
behaviour (unbiasedness, consistent error
matrix evaluation) of the technique. Moreover a comparison with the unbinned
extended likelihood results shows that the Modified Observable
estimators are practically reaching the maximum sensitivity. 
These two methods are completely 
equivalent at the ``perfect'' detector case (tables 1 and 2). 
A deterioration of the sensitivity (up to 20\%) when dealing with
realistic detectors is due  to
the imperfect  resolution of the measuring apparatus.\\
A comparison \cite{our183} between the sensitivity of 
several multiparametric TGC estimators, which include detector effects, 
shows that the Modified Observables are equivalent to the Iterative Optimal Variables
and Multidimensional Clustering techniques \cite{oonew}
whilst outperform classical methods of one or two dimensional binned likelihood fits.

\pagebreak

\pagebreak
\newpage

\begin{table}[tabsyst]
\begin{center}
\begin{tabular}{|c|c|c|c||} \hline \hline
  \em Method     &\multicolumn{3}{c|}
{   $\lambda_\gamma$ - $\Delta g_{1}^z$ } \\ \hline
\hline
       & $\sigma_{\lambda_\gamma}$  & $\sigma_{\Delta g_{1}^z}$  & $\rho$ \\\hline
``Perfect'' Extended Likelihood  & 0.21 $\pm$ 0.01 & 0.20 $\pm$ 0.01 & -0.73 $\pm$ 0.06\\ \hline
``Perfect'' Modified Observables    & 0.22 $\pm$ 0.01 & 0.21 $\pm$ 0.01 & -0.74 $\pm$ 0.06\\ \hline
Modified Observables    & 0.25 $\pm$ 0.01 & 0.23 $\pm$ 0.01 & -0.74 $\pm$ 0.06\\ \hline
\end{tabular} 
\end{center} 
\caption{Comparison of the statistical properties of
the technique proposed in this paper with the unbinned extended
likelihood estimations.}
{\label{tab1}} 
\end{table}

\begin{table}[tabsyst]
\begin{center}
\begin{tabular}{|c|c|c|c||}  \hline \hline
\em Method      &\multicolumn{3}{c|}{ $\Delta k_\gamma$ - $\Delta g_{1}^z $ } \\ \hline \hline
      & $\sigma_{\Delta k_\gamma}$ & $\sigma_{\Delta g_{1}^z} $ & $\rho$ \\ \hline
``Perfect'' Extended Likelihood  & 0.35  $\pm$ 0.03  & 0.14 $\pm$ 0.01 & -0.22 $\pm$ 0.08\\ \hline
``Perfect'' Modified Observables   & 0.38  $\pm$ 0.03  & 0.13 $\pm$ 0.01 & -0.25 $\pm$ 0.09\\ \hline
Modified Observables  & 0.44  $\pm$ 0.03  & 0.15 $\pm$ 0.01 & -0.28 $\pm$ 0.10\\ \hline
\end{tabular} 
\end{center} 
\caption{Comparison of the statistical properties of
the techniques proposed in this paper with the unbinned extended
likelihood estimations.}
{\label{tab2}} 
\end{table}

\begin{figure}[figure1]
\centerline{\epsfig{file=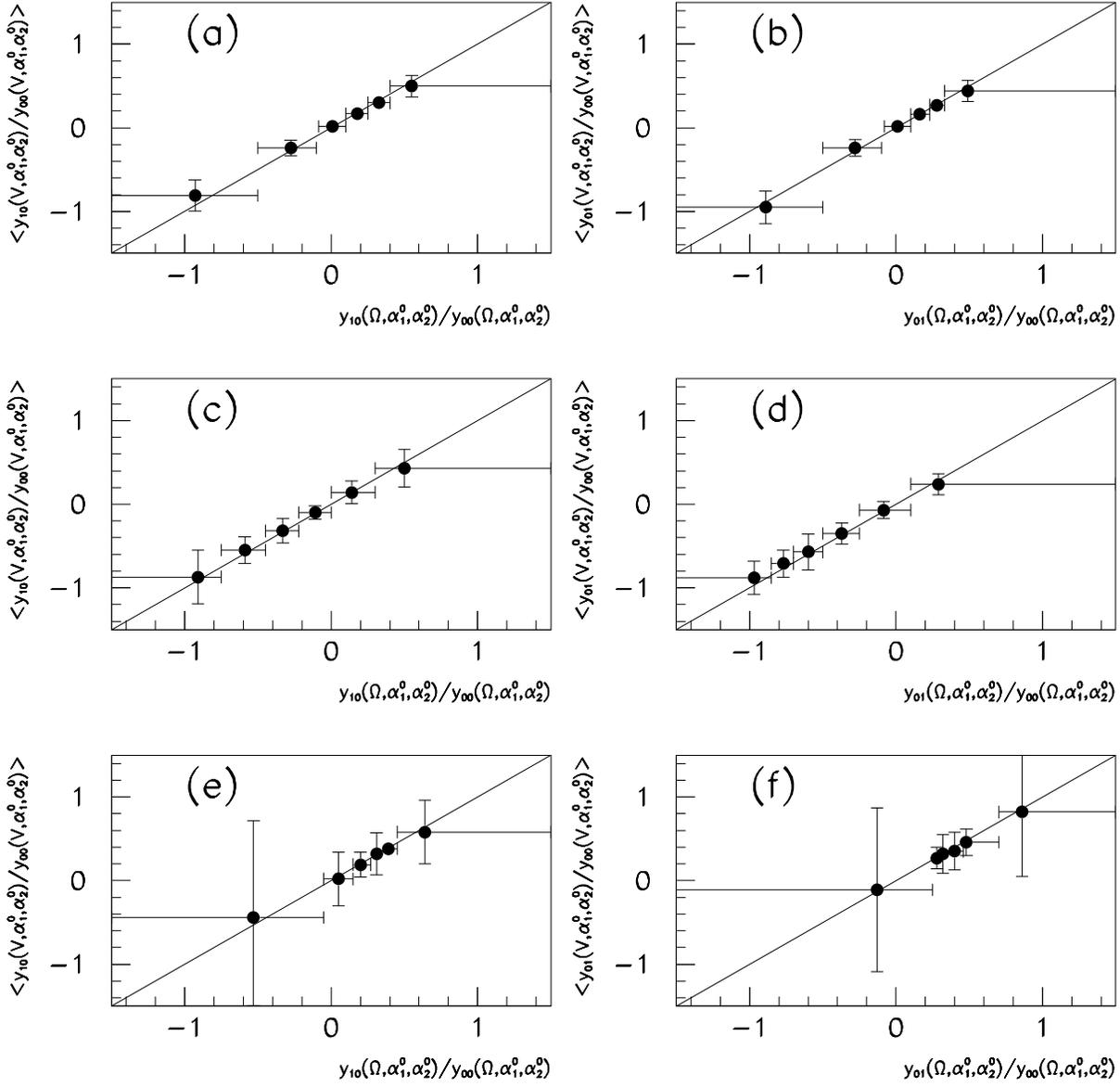,height=18cm}}
\caption
{
The mean values (see text) of the quantities 
$\frac {y_{10}(\vec V;\alpha_1^0,\alpha_2^0)}
       {y_{00}(\vec V;\alpha_1^0,\alpha_2^0)}$ 
[(a),(c),(e)] and
$\frac {y_{01}(\vec V;\alpha_1^0,\alpha_2^0)}
       {y_{00}(\vec V;\alpha_1^0,\alpha_2^0)}$
[(b),(d),(f)] as functions of the 
$\frac {y_{10}(\vec \Omega;\alpha_1^0,\alpha_2^0)}
       {y_{00}(\vec \Omega;\alpha_1^0,\alpha_2^0)}$ 
and
$\frac {y_{01}(\vec \Omega;\alpha_1^0,\alpha_2^0)}
       {y_{00}(\vec \Omega;\alpha_1^0,\alpha_2^0)}$
respectively.
These mean values correspond to kinematical vectors produced with couplings:
}
{
$\alpha_1 (\equiv \Delta g_{1}^{z})=0$,  
$\alpha_2 (\equiv \lambda_{\gamma})=0$ in (a) and (b)
}\\
{
$\alpha_1 (\equiv \Delta g_{1}^{z})=0$,
$\alpha_2 (\equiv \lambda_{\gamma})=-1$ in (c) and (d)
}\\
{
$\alpha_1 (\equiv \Delta g_{1}^{z})=0$, 
$\alpha_2 (\equiv \lambda_{\gamma})=+1$ in (e) and (f)
}

{\label{compare}}
\end{figure}


\begin{figure}[figure2]
\centerline{\epsfig{file=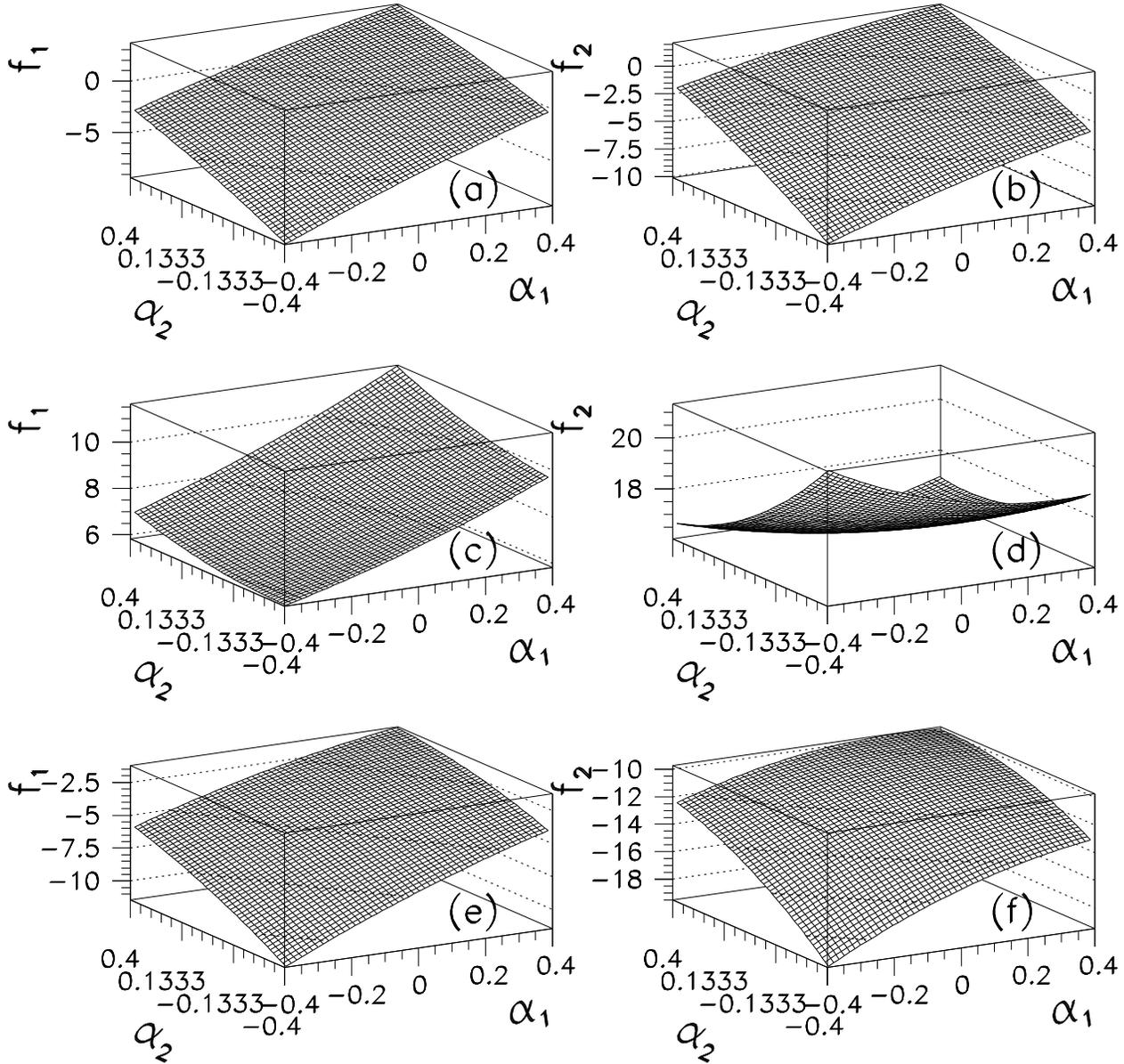,height=18cm}}
\caption
{
The $f_{1}(\alpha_1,\alpha_2;\alpha_1^0,\alpha_2^0)$,
$f_{2}(\alpha_1,\alpha_2;\alpha_1^0,\alpha_2^0)$
calibration surfaces for several expansion points
$\alpha_1^0 (\equiv \Delta g_{1}^{z})=0$ , 
$\alpha_2^0 (\equiv \lambda_{\gamma})=0$ in (a) and (b),
$\alpha_1^0 (\equiv \Delta g_{1}^{z})=0$, 
$\alpha_2^0 (\equiv \lambda_{\gamma})=+1$ in (c) and (d),
$\alpha_1^0 (\equiv \Delta g_{1}^{z})=0$, 
$\alpha_2^0 (\equiv \lambda_{\gamma})=-1$ in (e) and (f).
}
{\label{surfaces}}
\end{figure}

\begin{figure}[figure3]
\centerline{\epsfig{file=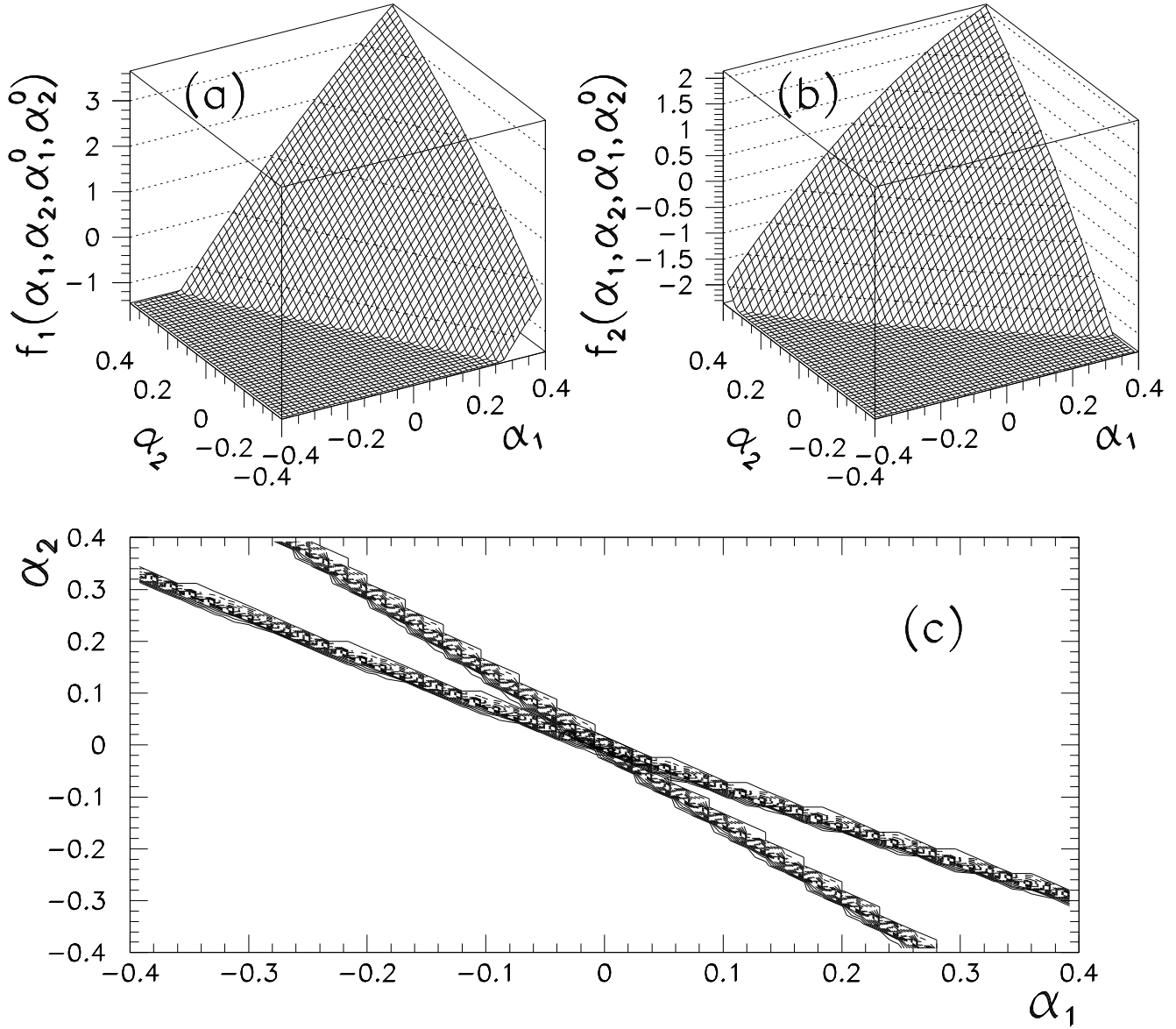,height=18cm}}
\caption
{
The calibration surfaces 
$f_{1}(\alpha_1,\alpha_2;\alpha_1^0,\alpha_2^0)$,
$f_{2}(\alpha_1,\alpha_2;\alpha_1^0,\alpha_2^0)$ as functions of the 
couplings $\alpha_1 = \Delta g_{1}^{z}$,
$\alpha_2 =  \lambda_{\gamma}$ at the expansion point
\{$\alpha_1^0=0$, $\alpha_2^0=0$\}. The horizontal shadowed planes correspond
to the experimental measurements. The two lines representing the intersection
of the calibration surfaces with the measured values are shown in (c).
}
{\label{surf2_00}}
\end{figure}

\begin{figure}[figure4]
\centerline{\epsfig{file=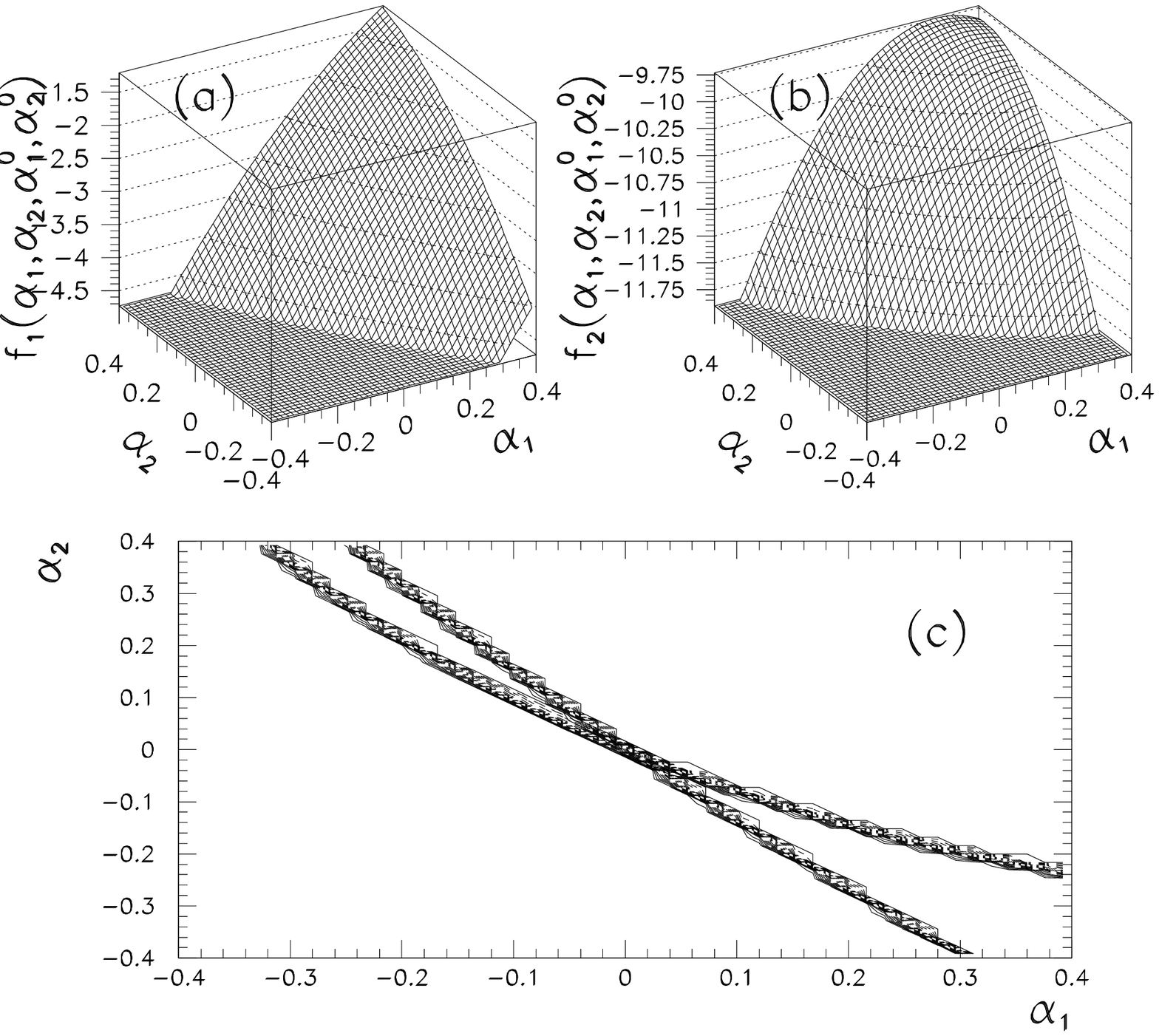,height=10cm,width=12cm}}
\centerline{\epsfig{file=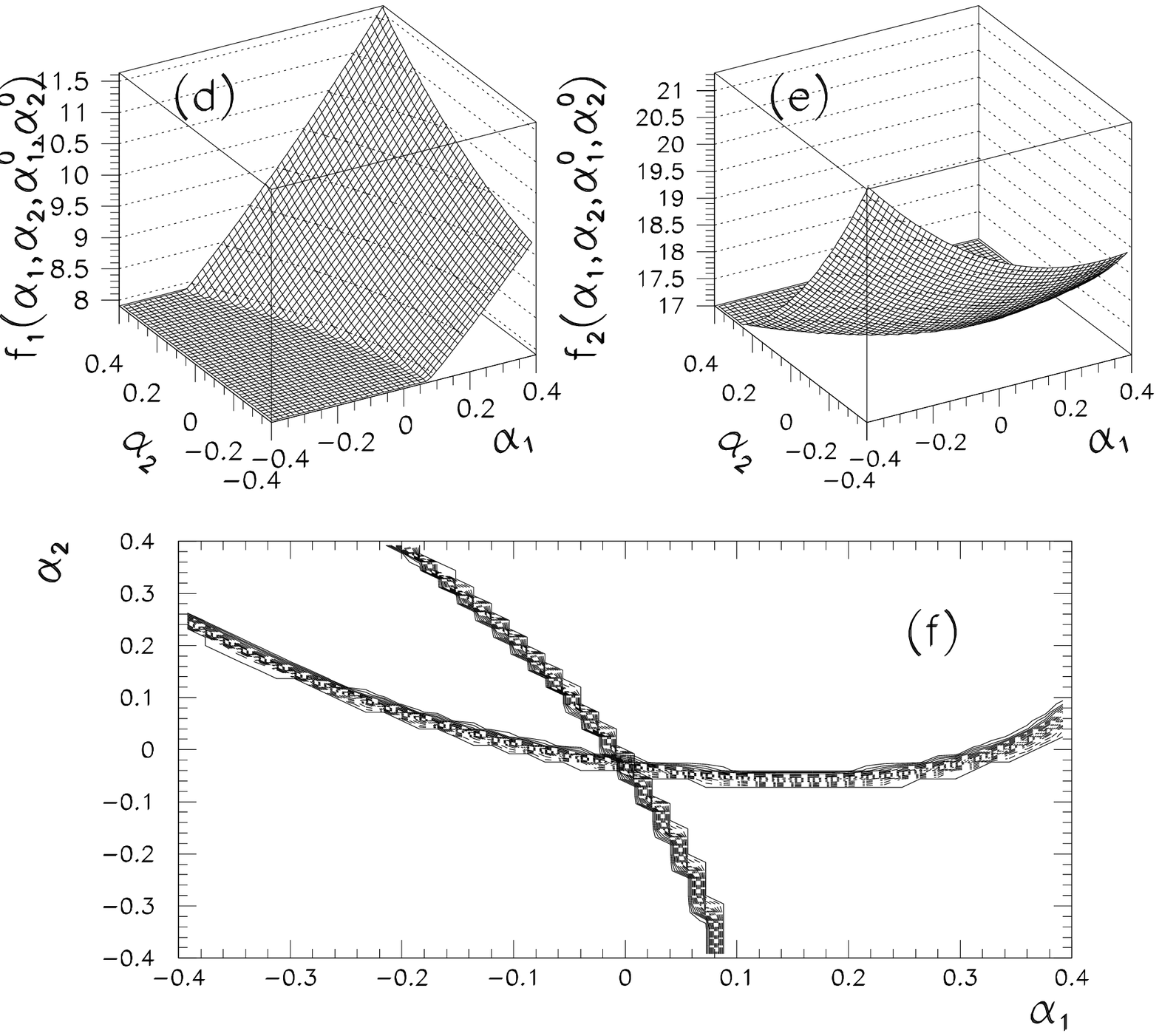,height=10cm,width=12cm}}
\caption
{
The calibration surfaces 
$f_{1}(\alpha_1,\alpha_2;\alpha_1^0,\alpha_2^0)$,
$f_{2}(\alpha_1,\alpha_2;\alpha_1^0,\alpha_2^0)$ as functions of the 
couplings $\alpha_1 = \Delta g_{1}^{z}$,
$\alpha_2 =  \lambda_{\gamma}$ at the expansion points 
\{$\alpha_1^0=0$, $\alpha_2^0=-1$\} in [(a),(b)] and 
\{$\alpha_1^0=0$, $\alpha_2^0=+1$\} in [(d),(e)]. 
The horizontal shadowed planes correspond to the experimental measurements. 
The two lines representing the intersection
of the calibration surfaces with the measured values are shown in (c) and (f) for the 
two pairs of expansion points, respectively.}
{\label{surf2_0m10p1}}
\end{figure}

\begin{figure}[figure5]
\centerline{\epsfig{file=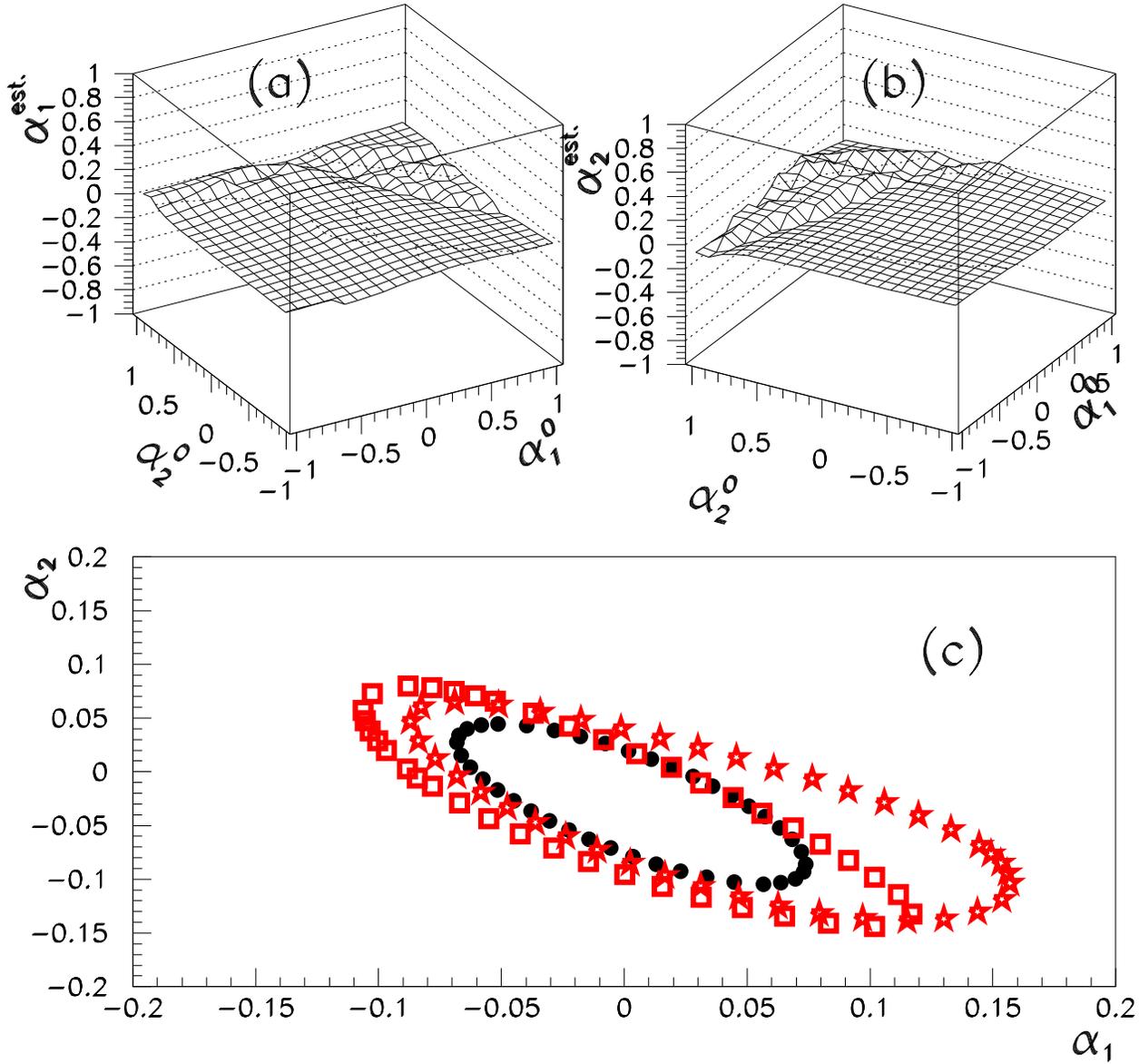,height=18cm}}
\caption
{
In [(a) and (b)] the estimated $\alpha_1 = \Delta g_{1}^{z}$ and
$\alpha_2 = \lambda_{\gamma}$ coupling values as functions of the expansion
points are shown. In (c) the evaluated 70\% C.L. contours for 
$\alpha_1^0=0, \alpha_2^0=0$ (solid points), 
$\alpha_1^0=0.2,\alpha_2^0=-0.6$ (squares) and 
$\alpha_1^0=-0.2,\alpha_2^0=0.6$ (stars) are compared.  
}
{\label{converge}}
\end{figure}

\begin{figure}[figure6]
\centerline{\epsfig{file=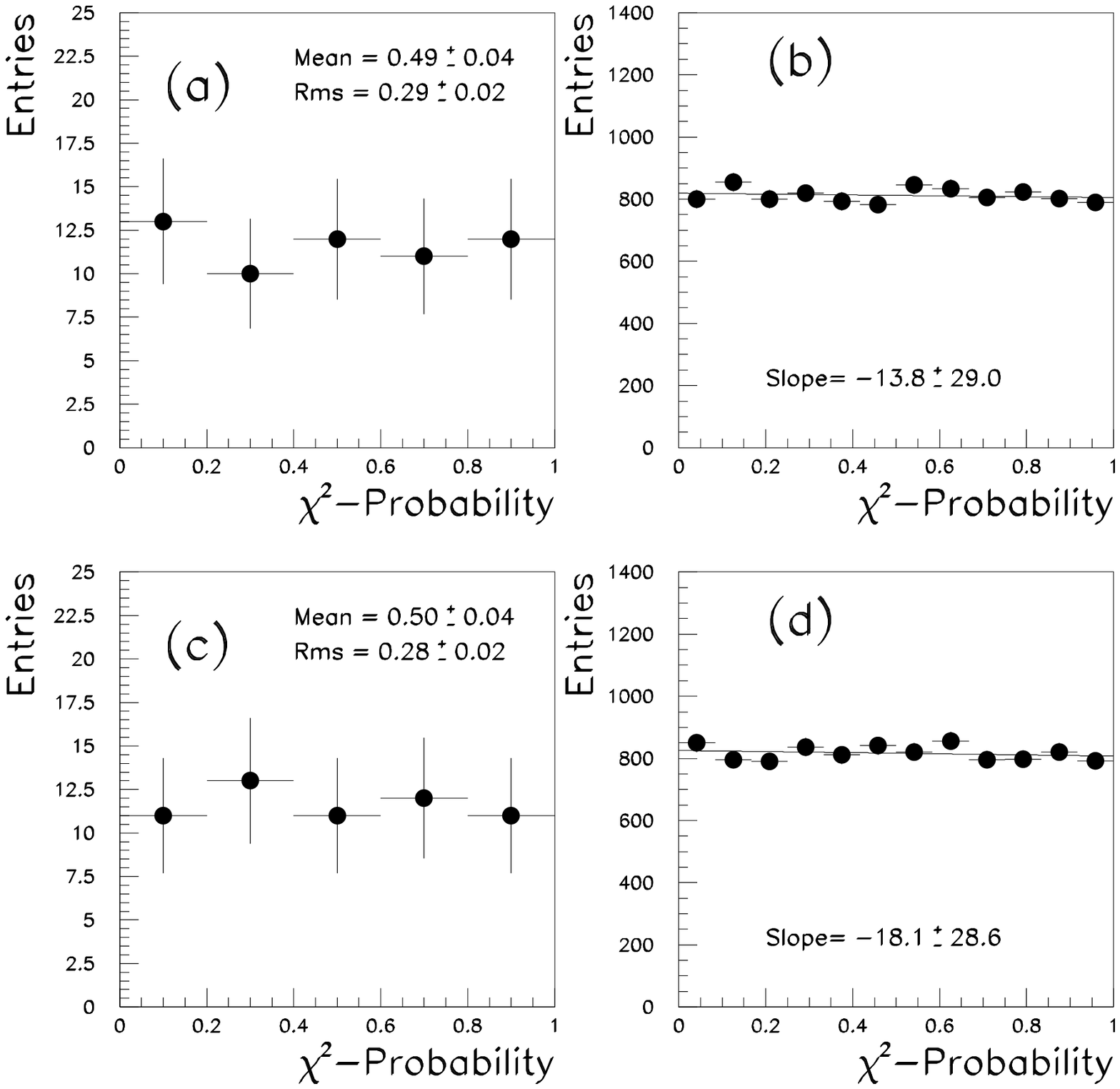,height=10cm,width=15cm}}
\centerline{\epsfig{file=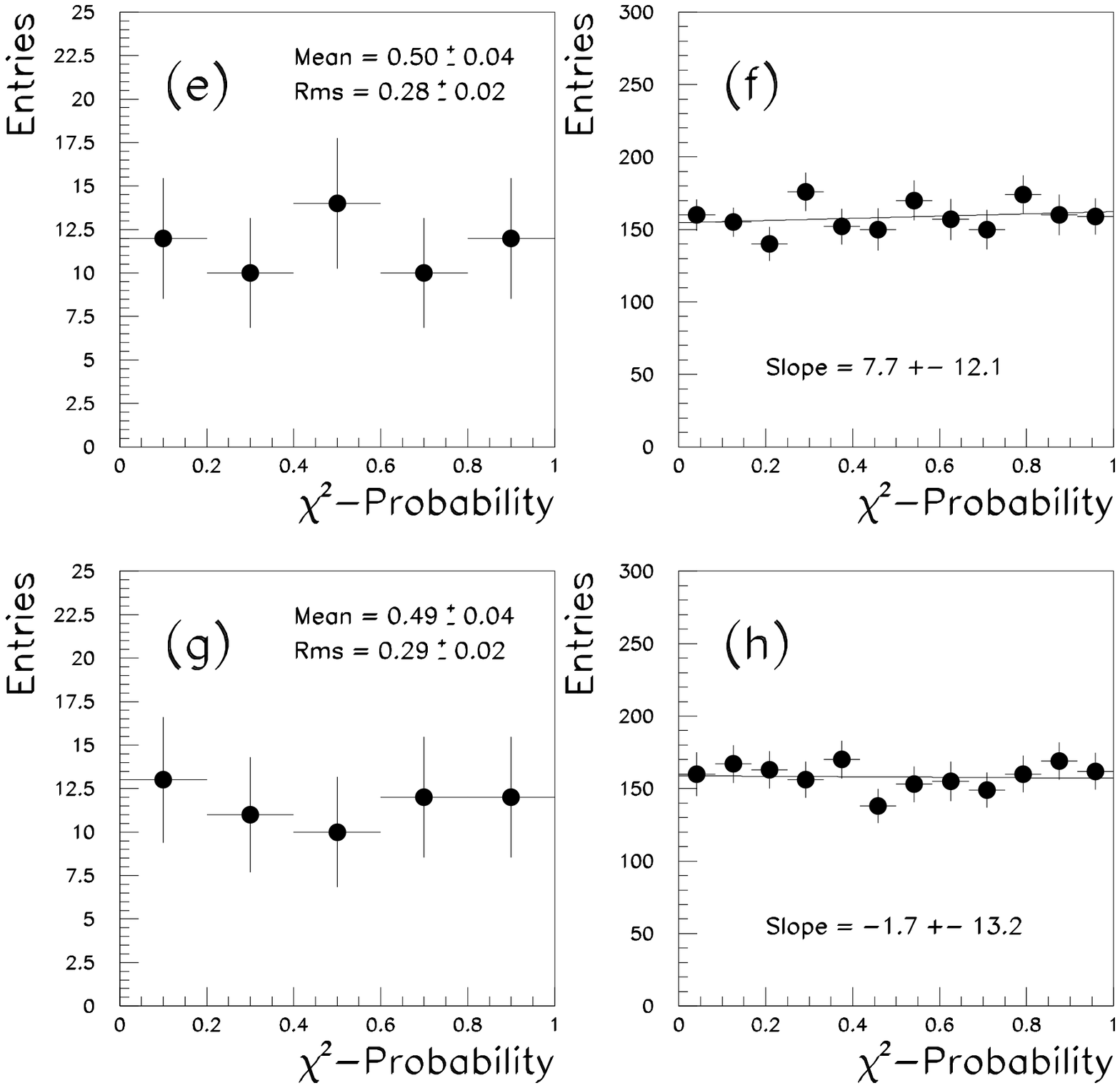,height=10cm,width=15cm}}
\caption
{
The distributions of $\chi^2$ (n.d.f.=2)  probabilities in obtaining 
$\lambda$ [(a),(b),(e),(f)] and
$\delta$  [(c),(d),(g),(h)] values in 
$\{\lambda_{\gamma}$,$\Delta g_{1}^{z}\}$ [(a),(b),(c),(d)] and 
$\{\Delta \kappa_{\gamma}$,$\Delta g_{1}^{z}\}$ 
[(e),(f),(g),(h)]
estimations by the Modified Observables technique.
The lines with slopes consistent with zero in (b),(d),(f) and (h) 
are first degree polynomial fits
to the bootstrap results.
}
{\label{chi}}
\end{figure}

\begin{figure}[figure7]
\centerline{\epsfig{file=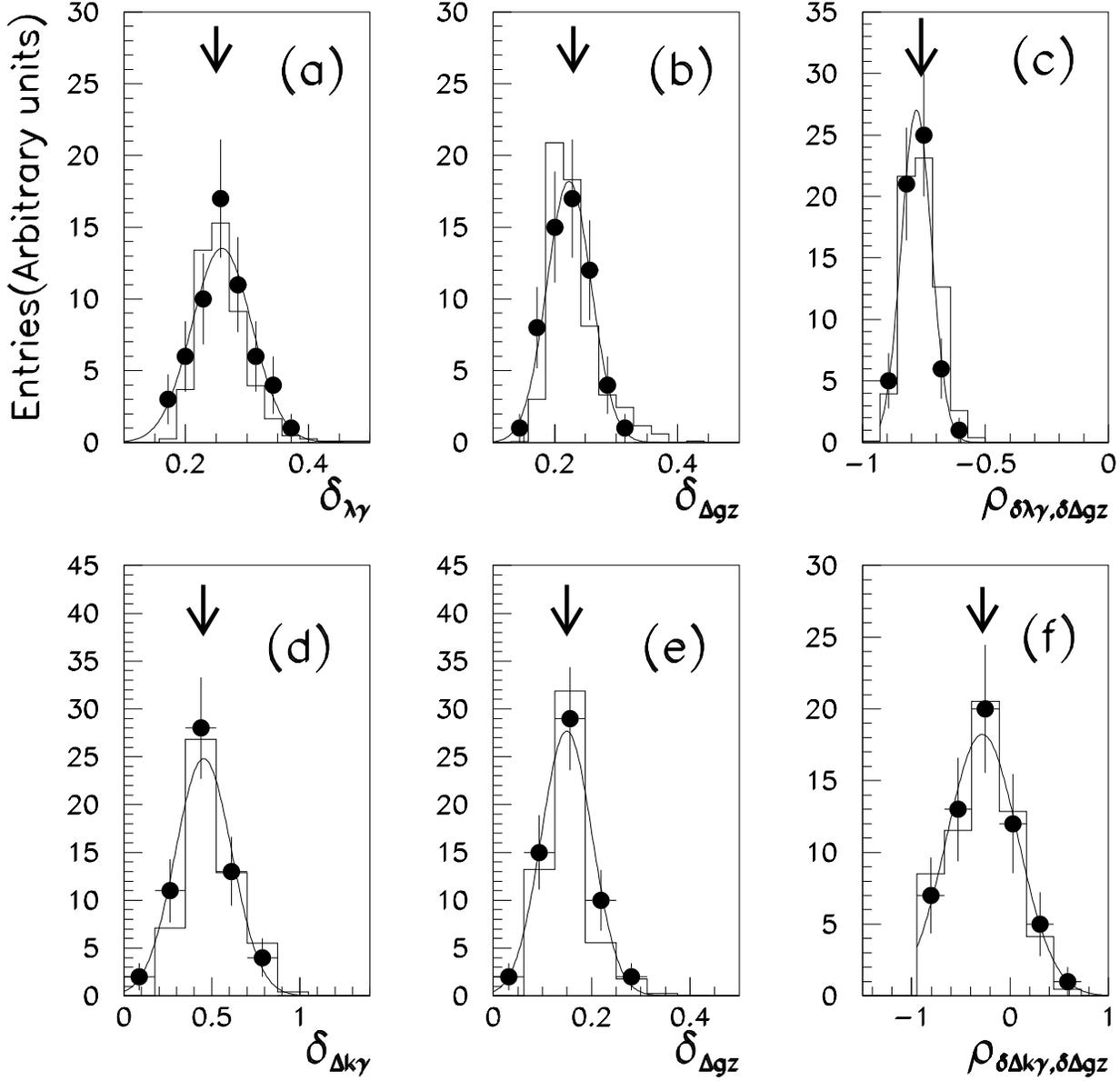,height=18cm}}
\caption
{
Confidence interval estimations with the Modified Observables technique.
The distributions of errors [(a),(b),(d),(e)] and 
correlations [(c),(f)] in estimating the 
$\{\lambda_{\gamma},\Delta g_{1}^{z}\}$ [(a),(b),(c)] and
$\{\Delta \kappa_{\gamma},\Delta g_{1}^{z}\}$ [(d),(e),(f)] pair of couplings.
The data points correspond to the 60 independent data sets whilst the 
histograms to the bootstrap results.
The arrows indicate the average sensitivities summarized in Tables 1 
and Table 2.
}
{\label{errors}}
\end{figure}

\begin{figure}[figure8]
\centerline{\epsfig{file=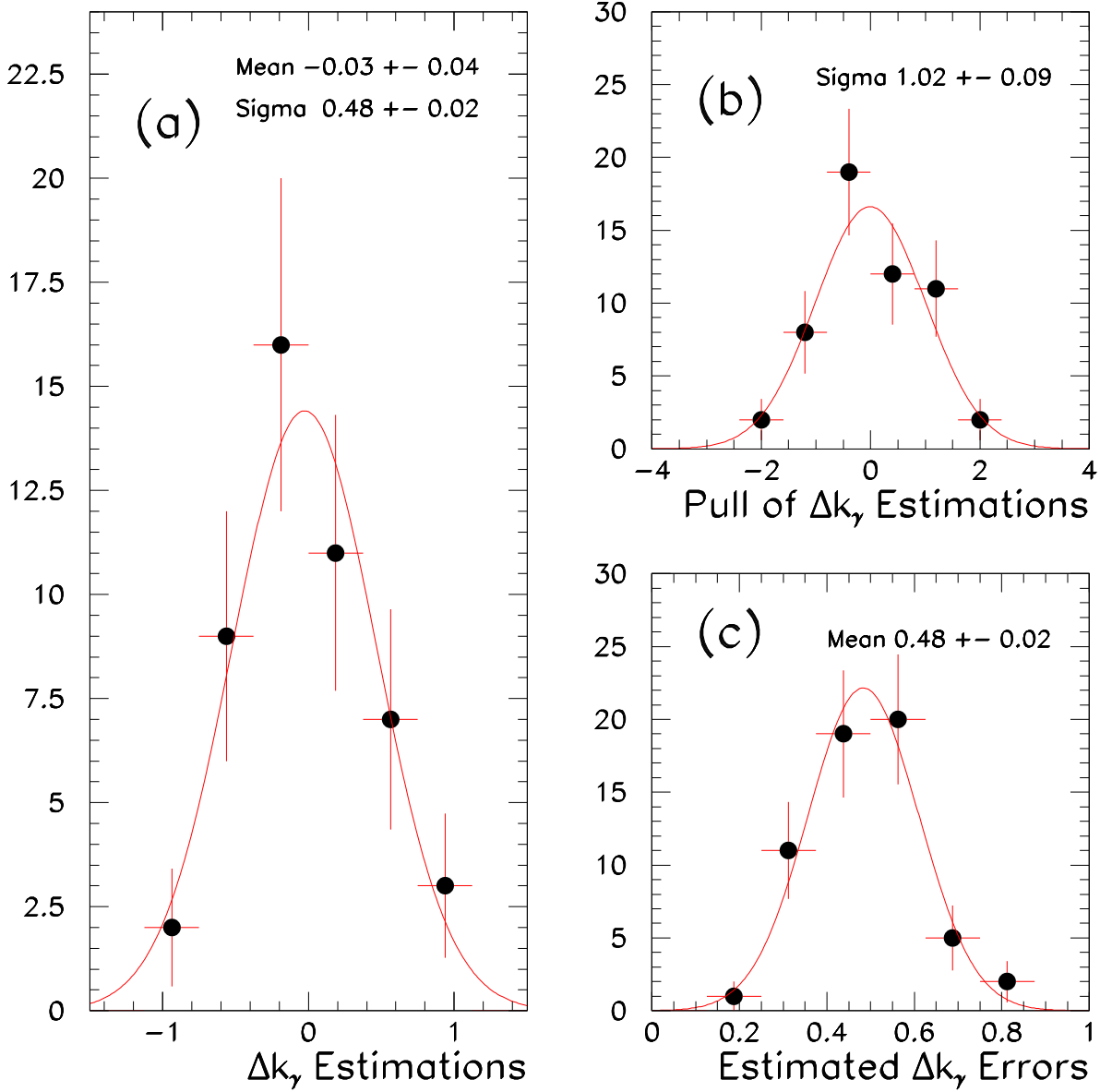,height=18cm}}
\caption
{
The sampling, pull and error distribution of $\Delta \kappa_{\gamma}$ 
estimation with the Modified Observable technique.
}
{\label{dkg}}
\end{figure}



\end{document}